\documentclass[preprint,showpacs] {revtex4-1}
\usepackage{amsmath,graphicx,amsfonts}

\usepackage{times}
\usepackage[colorlinks=true,linkcolor=blue,urlcolor=blue,citecolor=blue,breaklinks=true]{hyperref}

\usepackage[utf8]{inputenc}
\DeclareUnicodeCharacter{03BC}{\mbox{$\mu$}}
\DeclareUnicodeCharacter{2009}{ }

\def \dif {{d}}
\def\wt{\widetilde}
\def\vep{\varepsilon}
\def\eq#1{Eq.~(\ref{#1})}
\def\la{\langle}
\def\ra{\rangle}

\def\vep{\varepsilon}

\def\wt{\widetilde}

\newcommand{\be}{\begin{equation}}
\newcommand{\ee}{\end{equation}}
\newcommand{\bea}{\begin{eqnarray}}
\newcommand{\eea}{\end{eqnarray}}
\newcommand{\ba}{\begin{array}}
\newcommand{\ea}{\end{array}}
\newcommand{\bal}{\begin{align}}
\newcommand{\eal}{\end{align}}
\newcommand{\nn}{\nonumber\\}

\newcommand{\bem}{\begin{multline}}
\newcommand{\eem}{\end{multline}}

\def\nn{\hat n}
\def\tt{\hat\theta}

\def\da{\dagger}

\date{\today}
\begin{document}
	
\title{The Casimir-like effect in a one-dimensional Bose gas}
\author{Benjamin Reichert}
\author{Zoran Ristivojevic}
\author{Aleksandra Petkovi\'{c}}
\affiliation{Laboratoire de Physique Th\'{e}orique, Universit\'{e} de Toulouse, CNRS, UPS, 31062 Toulouse, France}

\begin{abstract}
The electromagnetic Casimir effect manifests as the interaction between uncharged conducting objects that are placed in a vacuum. More generally, the Casimir-like effect denotes an induced  interaction between external bodies  in a fluctuating medium. We study the Casimir-like interaction between two impurities embedded in a weakly interacting one-dimensional Bose gas.  We develop a theory based on the Gross-Pitaevskii equation that accounts for the effect of quantum fluctuations. At small separations, the induced interaction between the impurities decays exponentially with the distance. This is a classical result that can be understood using the mean-field Gross-Pitaevskii equation.  We find that at larger distances, the induced interaction crosses over into a power law dependence due to the quantum fluctuations. We obtain an analytic expression for the interaction that interpolates between the two limiting behaviors. The obtained result does not require any regularization.

\end{abstract}

\maketitle

\section {Introduction} In the most basic formulation, the electromagnetic Casimir effect is a macroscopic quantum phenomenon that represents the attraction between two large uncharged conducting plates that are placed in a vacuum \cite{casimir1948}. Despite early pioneering experiments, the accurate experimental confirmation of the Casimir prediction was achieved relatively recently \cite{klimchitskaya_casimir_2009}. Nowadays, the Casimir effect has become a multidisciplinary subject at the forefronts of many fields of modern physics, which includes gravitation, string theory, nanotechnology, and Bose-Einstein condensates \cite{Bordag2009}. 

The Casimir-like effect is a fluctuation-induced phenomenon. It denotes the interaction between external objects that are immersed in a fluctuating medium. Such objects impose constraints on otherwise free quantum fluctuations. As a result, the ground state energy of the system becomes modified giving rise to the Casimir-like interaction. It depends on the details of the correlations of fluctuations in the medium, which determine the interaction law \cite{kardar_``friction_1999,Bordag2009}. In the case of long-range correlations, the Casimir-like interaction is also expected to have a long-range nature.

The induced interaction between foreign particles, or impurities, embedded in a quantum liquid was studied in numerous works \cite{bardeen_effective_1967,bijlsma_phonon_2000,roberts_casimir-like_2005,klein_interaction_2005,recati_casimir_2005,fuchs_oscillating_2007,wachter_indirect_2007,schecter_phonon-mediated_2014,marino_casimir_2017,dehkharghani_coalescence_2018,pavlov_phonon-mediated_2018}. In particular, the one-dimensional liquid of repulsively interacting bosons was considered in Refs.~\cite{klein_interaction_2005,recati_casimir_2005,yu_casimir_2009,schecter_phonon-mediated_2014,dehkharghani_coalescence_2018}. In such a superfluid medium, a short-range attraction that decays exponentially beyond the healing length $\xi$ was found in Refs.~\cite{klein_interaction_2005,recati_casimir_2005,dehkharghani_coalescence_2018}. However, although the one-dimensional superfluid possesses long-range correlations, the long-range Casimir-like interaction was not found in these works. Only recently, \citet{schecter_phonon-mediated_2014} have found the long-range interaction. At large distances $\ell$ between the impurities, it scales as $1/\ell^3$. In the opposite limit $\ell \to 0$, they found the contact interaction. 

We note that the divergence of these results can be partly explained by the employment of different approaches having their own shortcomings. The exponential decay \cite{klein_interaction_2005,recati_casimir_2005,dehkharghani_coalescence_2018} was obtained using the microscopic approach based on the Gross-Pitaevskii equation. Such mean-field  method does not account for the effect of quantum fluctuations, which is necessary to describe the long-range Casimir-like interaction. On the other hand, Ref.~\cite{schecter_phonon-mediated_2014} uses a mobile impurity formalism within the framework of the Luttinger liquid theory. Such phenomenological approach has its own complementary limitations.  It properly describes the low energy physics where the quasiparticle spectrum can be approximated by the linear one. Thus \citet{schecter_phonon-mediated_2014} were able to describe the Casimir-like interaction only at very large distances, $\ell\gg \xi$. The contact interaction between impurities obtained in Ref.~\cite{schecter_phonon-mediated_2014} is an artifact of the above-mentioned limitation.

An impurity that interacts repulsively with the particles of the liquid produces the depletion of the liquid density. Such effect occurs locally; its spatial extent is set by the healing length $\xi$. In the presence of two impurities, it becomes energetically more favorable that the two depletion regions overlap. Hence there is an induced attraction between the impurities at small distances, $\ell\lesssim \xi$. At separations above $\xi$, the two depletion clouds practically do not overlap and thus do not interact within such a classical picture. However,  as a consequence of (quasi-)long-range correlations of fluctuations in the liquid, the two depletion regions indeed feel each other. This mechanism produces the long-range Casimir-like interaction between the impurities. Such contribution can be seen as a quantum fluctuation correction to the classical contribution that prevails at small distances. The described scenario is common in other physical problems. For example, the domain walls (solitons) in the system of atoms adsorbed on a periodic substrate interact exponentially at small separations \cite{pokrovsky_1986}. Due to entropic effects, this dependence crosses over  into a long-range power law once one accounts for the effect of thermal fluctuations  \cite{pokrovsky_1986}. We also note that the  electromagnetic Casimir-Polder interaction \cite{casimir_influence_1948} between neutral atoms, which are usually the realization of impurities in Bose liquids, decays faster than the induced interaction studied here, as we discuss below.

\section{Results}

In this work we consider the Casimir-like effect in a one-dimensional Bose liquid. We develop the consistent microscopic approach that overcomes the above explained shortcomings present in other studies \cite{klein_interaction_2005,recati_casimir_2005,dehkharghani_coalescence_2018,schecter_phonon-mediated_2014}. Within our description we are able to simultaneously treat the nonlinear spectrum of quasiparticles and the effect of quantum fluctuations. 
We find the analytical expression for the effective impurity interaction 
\begin{align}\label{eq:Final}
U(\ell)={}&-\frac{G^2 m}{\hbar^2 \sqrt{\gamma}}e^{-2\ell /\xi}\left\{1+\frac{\sqrt{\gamma}}{2\pi}\left[1-\frac{2\ell}{\xi}+J\left(\frac{\ell}{\xi}\right)\right]\right\}+\frac{G^2 m\xi^3}{64\pi\hbar^2\ell^3}\int_0^\infty  \frac{dx \;x^2\sin x}{(1+\frac{\xi^2 x^2}{8\ell^2}) \sqrt{1+\frac{\xi^2 x^2}{16\ell^2}}}
\end{align}
that is valid at an arbitrary distance $\ell$ between the impurities.
In Eq.~(\ref{eq:Final}), $m$ is the mass of bosonic particles, $G$ denotes the coupling strength of the impurities to the bosonic subsystem, the small dimensionless parameter $\gamma\ll 1$ characterizes the interaction between bosons, while $\xi=1/n\sqrt\gamma$ denotes the healing length at weak interaction \cite{pitaevskii_bose-einstein_2003}. By $n$ is denoted the density of bosons. In Eq.~(\ref{eq:Final}), 
\begin{align}
J(z)=\int_0^\infty dx\left[ \frac{2}{x\sqrt{4+x^2}}-\frac{8e^{-z(\sqrt{4+x^2}-2)}\cos(z x)}{x(2+x^2)(4+x^2)}\right]
\end{align}
is the monotonically increasing function. It has the limiting behavior $J(0)=\ln 4$ while at large $z$, $J(z)$ grows logarithmically.

\begin{figure}
	\centering
	\includegraphics[scale=0.6]{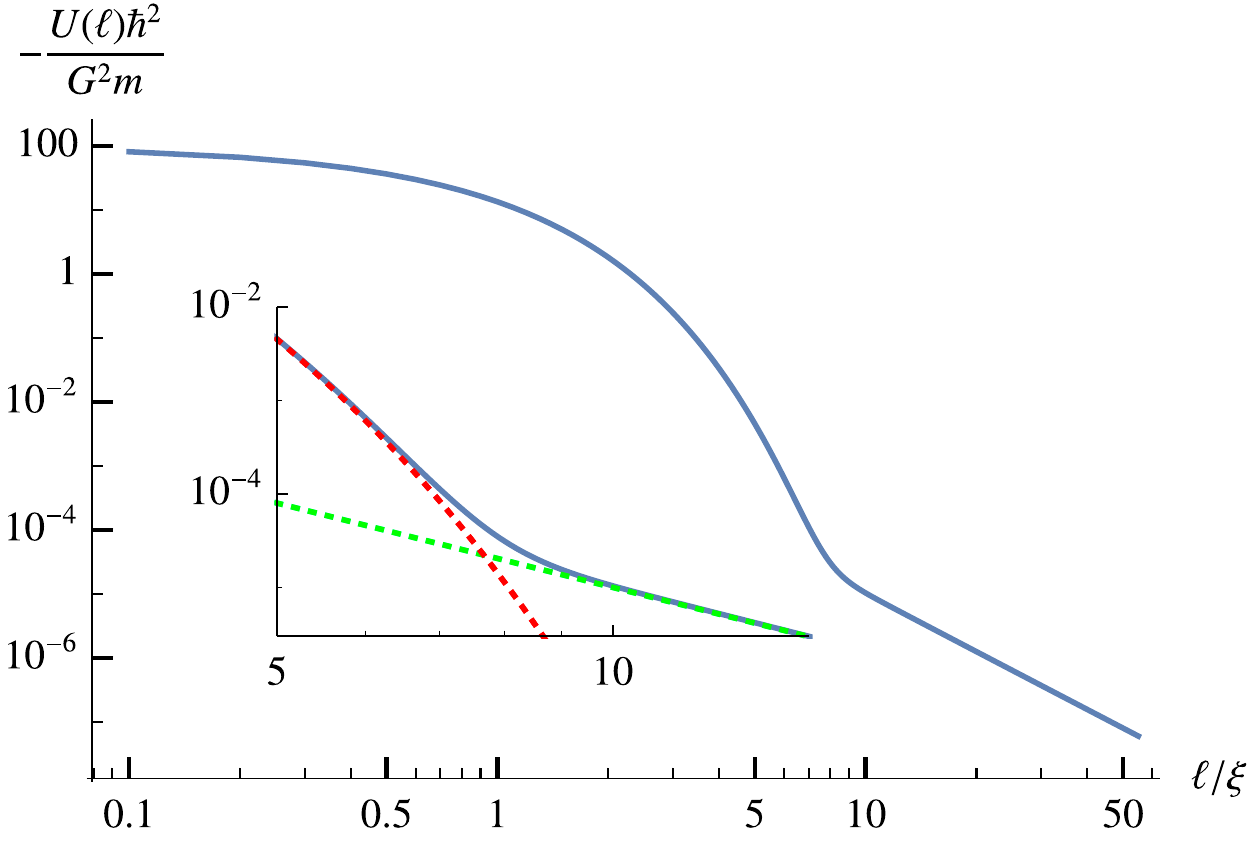}
	\caption{The induced interaction between two impurities as a function of their distance for $\gamma=10^{-4}$. The inset is a zoom around the crossover, where by the dashed lines are shown the two asymptotes given by Eqs.~(\ref{Uexp}) and (\ref{Uinf}).}
	\label{fig:plot}
\end{figure}

At small distances, $\ell\lesssim \xi$, the dominant part of the interaction (\ref{eq:Final}) is given by the exponential
\begin{align}\label{Uexp}
U(\ell)=-\frac{G^2 m}{\hbar^2 \sqrt{\gamma}}e^{-2\ell /\xi}.
\end{align}
This is a classical result that could be understood using the mean-field Gross-Pitaevskii equation \cite{klein_interaction_2005,recati_casimir_2005,dehkharghani_coalescence_2018}. Our expression (\ref{eq:Final})  contains the quantum correction to Eq.~(\ref{Uexp}). While the classical result behaves as $\gamma^{-1/2}$, the leading quantum correction is proportional to $\gamma^0$ and one would naively think that it is always subleading. However, the exponential contribution (\ref{Uexp}) becomes negligible at large distances and  therefore the quantum correction prevails in this regime. At large distances, $\ell\gg\xi$, the last term in Eq.~(\ref{eq:Final}) dominates and yields 
\begin{align}\label{Uinf}
U(\ell)=-\frac{G^2 m \xi^3}{32\pi\hbar^2 \ell^3} \left[1+\frac{15\xi^2}{8\ell^2}+\mathcal O\left( \frac{\xi^4}{\ell^4}\right)\right].
\end{align}
We have thus obtained the long-range Casimir-like attraction that scales with the third power of the inverse distance.  The leading order term in Eq.~(\ref{Uinf}) is in full agreement with Ref.~\cite{schecter_phonon-mediated_2014}.  However, our result (\ref{eq:Final}) describes the effective interaction in the whole crossover region between the two limiting cases of small and large $\ell$, as shown in  Fig.~\ref{fig:plot}. Equation (\ref{eq:Final}) shows that there are no new intermediate regimes in the induced interaction. We eventually note that unlike some other approaches and setups where the Casimir-like effect was studied, our formalism properly accounts for the physics essentially at all energy scales and hence the final formula (\ref{eq:Final}) does not require regularization. Equation (\ref{eq:Final}) is our main result. In the remainder of the paper we derive it. 

\section{Methods} We study the system of one-dimensional bosons with contact repulsion in the presence of two impurities. It is described by the Hamiltonian
\begin{align}
		H=&\int dx\left(-\hat\Psi^\da\dfrac{\hbar^2\partial_x^2}{2m}\hat\Psi+\frac{g}{2}\hat\Psi^\da\hat\Psi^\da\hat\Psi\hat\Psi\right)\notag \\ &+G\left[\hat\Psi^\da(\ell/2,t)\hat \Psi(\ell/2,t)+\hat\Psi^\da(-\ell/2,t)\hat \Psi(-\ell/2,t)\right],
		\label{eq:H}
\end{align}
where $\hat\Psi^\da$ and $\hat\Psi$ are the bosonic single-particle operators that satisfy the standard commutation relation $[ \hat\Psi(x,t) , \hat\Psi^\da(x',t)]=\delta(x-x')$. The repulsion between the bosons has the strength $g$ which enters the dimensionless parameter $\gamma=m g/\hbar^2 n$.  The two impurities at positions $\pm \ell/2$ locally couple to the density of the Bose liquid with the strength $G$. At $G=0$, the Hamiltonian (\ref{eq:H}) corresponds to the integrable Lieb-Liniger model \cite{lieb_exact_1963I}. 

Our goal is to compute analytically the dependence of the ground-state energy of the Hamiltonian (\ref{eq:H}) on the impurity separation $\ell$ and hence find the induced interaction between the impurities. For convenience, in the following we use the dimensionless quantities. The length is measured in units of $\hbar/\sqrt{m\mu}$ and the time in units of $\hbar/\mu$, where $\mu$ denotes the chemical potential. We express the field operator as $\hat\Psi(x,t)=\sqrt{\mu/g}\,\hat\psi (X,T)\, e^{-iT}$, where $X$ and $T$ are the dimensionless coordinate and time, respectively. Then, the equation of motion becomes
\begin{align}
	i\partial_T \hat\psi=\left[ -\partial_X^2/2 -1+\hat\psi^\da\hat\psi+L\,\wt G\delta(X^2-L^2/4)\right] \hat\psi,
	\label{eq:eom}
\end{align}where $L$ is the dimensionless distance between impurities and $G= \hbar \sqrt{\mu/m} \,\wt G$.  
In the case of weakly interacting bosons, the field operator  can be expanded as \cite{pitaevskii_bose-einstein_2003,sykes_drag_2009}
\begin{gather}
		\hat\psi(X,T)=\psi_0(X)+\alpha\hat\psi_1(X,T)+\alpha^2\hat\psi_2(X,T)+\ldots, \label{eq:wf_exp}
\end{gather}
such that $[ \hat\psi(X,T) , \hat\psi^\da(X',T)]=\alpha^2 \delta(X-X')$. Here  $\alpha=(\gamma gn/\mu)^{1/4}\approx \gamma^{1/4}\ll1$ is a small parameter. The field $\psi_0(X)$ describes the time independent condensate wave function in the absence of  fluctuations, while  the operators $\hat\psi_{1,2}(X,T)$ account for the effects of quantum fluctuations. We point out that condensate does not exist in one dimension in the thermodynamic limit due to the strong effect of long-wavelength fluctuations. However, in finite-size systems,  the inverse system length provides an infrared cutoff. For (dimensionless) system length that satisfies $\ln \widetilde L_s\ll1/\sqrt{\gamma}$ \cite{petrov_low-dimensional_2004,sykes_drag_2009}, the fluctuation contributions in Eq.~(\ref{eq:wf_exp}) remain smaller than the condensate contribution and one can use the expansion (\ref{eq:wf_exp}). We notice that at weak interaction, $\gamma\ll 1$, the system size can be huge. In Appendix we present a complementary approach for the calculation of the induced interaction (\ref{eq:Final}). It is based on the density-phase representation of the field operator and does not rely on the expansion (\ref{eq:wf_exp}). It allows us to express $\psi_0=\sqrt{n_0}$, where $n_0$ is the mean-field density in the system, and $\hat\psi_{1,2}$ in terms of the fluctuation contributions to the density and the phase such that the equations for $\hat\psi_{0,1,2}$ given in the text below are valid.  The method presented in Appendix leads to  Eq.~(\ref{eq:Final}). Thus the result (\ref{eq:Final}) does not require the expansion (\ref{eq:wf_exp}) and the above-discussed infrared cutoff. 

We now solve Eq.~(\ref{eq:eom}) order by order in $\alpha$. 
The condensate wave function $\psi_0(X)$ is obtained by solving the equation of motion (\ref{eq:eom})  at  order $\alpha^0$. It reads as
\begin{align}\label{Psi_0}
\left[-\partial_X^2/2 +|\psi_0(X)|^2-1+L\,\wt G\delta(X^2-L^2/4)\right]\psi_0(X)=0
\end{align}
and is known as the Gross-Pitaevskii equation. In the case of a homogeneous Bose gas (i.e., without impurities), the condensate density is constant, $|\psi_0(X)|^2=1$. If weak repulsive impurities  are added into the Bose gas,  a  depletion of the condensate occurs in the vicinity of the impurities. However, far from the impurities such disturbance is not visible, $\lim_{X\to \pm \infty}|\psi_0(X)|= 1$. 
		 
We first find the solutions for $\psi_0(X)$ in the case $\wt G=0$ in the three regions: $X<-L/2$, $|X|<L/2$, and $X>L/2$. We then match them such that the overall solution is continuous $\psi_0(\pm L/2^+)=\psi_0(\pm L/2^-)$ but its derivative is  discontinuous: $\psi_0'(\pm L/2^+)=\psi_0'(\pm L/2^-)+2\wt G\psi_0(\pm L/2)$. Here the prime denotes the derivative with respect to $X$  and $f(X^\pm)=\lim_{\vep\to 0^+}f(X \pm \vep)$. Since the system is invariant with respect to the inversion, $X \to -X$, the density also possesses this symmetry. We find that the solution with vanishing supercurrent at infinity satisfies $
|\psi_0(X)|= \tanh(|X|-L/2+b)$ for $|X|\geq L/2$ and $|\psi_0(X)|=\sqrt{{2a}/({1+a})}\,\text{cd}\left(\sqrt{2} X/\sqrt{{1+a}} ;a\right)$ for $|X|<L/2$, while the argument of the complex wave function is a constant. Since the solution of the Gross-Pitaevskii equation is undetermined up to a phase factor, the argument of $\psi_0(X)$ is set to zero for simplicity.  By $\text{cd}(x,y)$ we denote the Jacobi elliptic function, while $a$ and $b$ are the parameters to be determined from constraints on the continuity of the function and the discontinuity of its derivative at the impurity positions. In the following we consider weakly coupled impurities, $\wt G\ll 1$. For $\wt G\to 0$, the wave function has to satisfy $|\psi_0(X)|^2=1$. The latter is achieved with the choice $b=\infty$ and $a=1$. The expansion in $\wt G$ of $a$ and $b$  around these values is given by $a=1-4 \wt G e^{-L}+\mathcal O\big(\wt G^{2}\big)$ and $b=-\ln\left[ \wt G(1+e^{-2L})/{4}\right]/2+\mathcal O\big(\wt G\big)$, respectively. To linear order in $\wt G$ we then obtain
	\begin{align}
		\psi_0(X)=1-
		\wt G\begin{cases}
		 e^{-2|X|} \cosh L,&|X|\geq L/2,\\
		 e^{-L}\cosh 2X, &|X|<L/2.	
		\end{cases}
		\label{eq:psi0}
	\end{align}
Equation (\ref{eq:psi0}) is the wave function of the condensate where the fluctuations are neglected. It leads to the exponential interaction (\ref{Uexp}), as we explain below.

Now we consider the effect of fluctuations represented by the field operator $\hat\psi_1$. Its equation of motion is obtained from \eq{eq:eom} at order $\alpha$ and reads as 
\begin{align}\label{eqPsi1}
		i\partial_T \hat\psi_1=\left(-{\partial^2_X}/{2}-1+2|\psi_0|^2\right)\hat\psi_1+\psi_0^2\hat\psi_1^\da +L\,\wt G\delta(X^2-L^2/4)\hat\psi_1.
\end{align} 
We seek for $\hat\psi_1(X,T)$ in the form of a superposition of excitations of energy $\epsilon_k$ \cite{pitaevskii_bose-einstein_2003}
\begin{align}
\hat\psi_1(X,T)=\sum_k N_k \left[u_k(X)b_k e^{-i\epsilon_k T}-v_k^*(X)b_k^\da e^{i \epsilon_k T}\right],
\label{eq:uv}
\end{align}
where $N_k$ is a normalization factor. The bosonic operators $b_k$ and $b_k^\da$ obey the standard commutation relation $[b_k,b_{k'}^\dagger]=\delta_{k,k'}$. We will first solve Eq.~(\ref{eqPsi1})  without the $\delta$-potential. We will then take them into account through the boundary conditions
$u_k(\pm L/2^+)=u_k(\pm L/2^-)$ and  $u_k'(\pm L/2^+)=u_k'(\pm L/2^-)+2\wt Gu_k(\pm L/2)$.  The same two constraints are to be imposed to $v_k$. Here the prime denotes the derivative with respect to $X$. 

Substituting \eq{eq:uv} into Eq.~(\ref{eqPsi1}) without the $\delta$-potential, we obtain a system of two coupled equations for $u_k$ and $v_k$ known as the Bogoliubov-de Gennes equations. The system can be simplified by introducing the functions $S(k,X)=u_k(X)+v_k(X)$ and $D(k,X)=u_k(X)-v_k(X)$. After some algebra we obtain:
\begin{align}
\epsilon_k^2 S(k,X)=
		\left[-{\partial^2_X }/2+ 3\psi_0^2(X)-1\right]\left[-{\partial^2_X }/2+\psi_0^2(X)-1\right]S(k,X)
\end{align} 
and $
\epsilon_k D(k,X)=\left[-{\partial^2_X }/2+\psi_0^2(X)-1\right]S(k,X)$. Once one determines the function $S$, the solution for $D$ follows directly from the latter equation. 

It is instructive to first solve the Bogoliubov-de Gennes equations in the case of a homogeneous condensate, $\psi_0(X)=1$. We assume a solution in the form $S(k,X)=e^{ikX}$ and obtain that $\epsilon_k$ satisfies the Bogoliubov dispersion relation $\epsilon_k^2=k^2+k^4/4$. For a fixed energy $\epsilon_k$ there are four different momenta, $k_{1,2}=\pm k$ and $k_{3,4}=\pm i\sqrt{4+k^2}$, where 
$k=\sqrt{2}\sqrt{\sqrt{1+\epsilon_k^2}-1}$.
The solutions with momenta  $k_{1,2}$ represent propagating waves, while the ones with $k_{3,4}$ show the exponential behavior. The general solution for $S$ in the case $\wt G=0$ is a linear combination of these four  solutions.
In the presence of the two impurities,  the background density $\psi_0^2(X)$ is  $X$-dependent [see \eq{eq:psi0}]. We assume a solution in the form $S(k,X)=f(k,X,\wt G)e^{ikX}$, where $f(k,X,\wt G\to 0)=1$, and find four linearly independent solutions. To linear order in $\tilde G$, the Bogoliubov dispersion remains unchanged and the solutions read as
\begin{widetext}
	\begin{align}
		S_n(k,X)=e^{ik_n X}\begin{cases}
			1-2\wt G \dfrac{[2+i k_n \text{sgn}(X)] e^{-2|X|} \cosh L}{4+k_n^2},  & |X|\geq L/2,\\
			1-2\wt G \dfrac{e^{-L}(2\cosh 2X-i k_n \sinh 2X)}{4+k_n^2},	&|X|<L/2,
		\end{cases}
	\label{eq:Sn}
	\end{align}
\end{widetext}
where $n\in \{1,2,3,4\}$.  Using Eq.~(\ref{eq:Sn}) and the Bogoliubov-de Gennes equation, one can easily generate the corresponding $D_n$ functions.

It remains to account for the  $\delta$-potential in Eq.~(\ref{eqPsi1}). The general solution for $S(k,X)$ is a linear combination of the four solutions (\ref{eq:Sn}). The coefficients will be determined from the boundary conditions. There are formally two scattering situations to be considered: one with an incoming wave (i) from $X=-\infty$ with a positive wave number, and the other with a wave (ii) from $X=+\infty$ with a negative wave number. Since the impurities are static, the two problems are related and one needs to analyze only one of them. For example, knowing the result $S(k,X)$ for the problem (i), the solution for the problem (ii) is  given by $S(k,-X)$. We can use this fact to define solutions for negative wave numbers as $S(-k,X)=S(k,-X)$, where $k$ is assumed to be positive. The general solution  for a positive wave number $k>0$ takes the form
\begin{widetext}
\begin{align}
		S(k,X)=
		\begin{cases}
			S_1(k,X)+l_1S_2(k,X)+l_2S_4(k,X), &X<-L/2,\\
			l_3S_1(k,X)+l_4S_2(k,X)+l_5S_3(k,X)+l_6S_4(k,X),&|X|\leq L/2,\\
			l_7S_1(k,X)+l_8S_3(k,X),&X>L/2,
			\label{eq:Ssetup}
		\end{cases} 
\end{align}
\end{widetext}
where we excluded the exponentially diverging solutions for $|X|>L/2$. 
However, there are "bound states" around impurity positions described by the exponential solutions ($S_3$ and $S_4$). 
The general solution for $D(k,X)$ satisfies also Eq.~(\ref{eq:Ssetup}) where the functions $S_n$ are replaced with $D_n$. 
In Eq.~(\ref{eq:Ssetup}), $l_1,\ldots ,l_8 $ are eight parameters to be determined by imposing the  continuity of the function $S(k,\pm L/2^{-})=S(k,\pm L/2^{+})$, by  the discontinuity of its derivative  $S'(k,\pm L/2^{+})-S'(k,\pm L/2^{-})=2\wt G S(k,\pm L/2^{+})$ at impurity positions and the equivalent conditions for $D(k,X)$. 
To linear order in $\wt G$, we find : $-l_1/(2\cos{kL})=(l_7-1)/2=l_3-1=-l_4e^{-i kL}=i\wt G{k}/({2+k^2})$. Also,  $l_2=l_8^*={4}i\wt Gk/[{(2+k^2)(4+k^2)}]\cos \left[(k+i \sqrt{4+k^2})L/2\right]$, and  $l_6^*=l_5=-2i\wt G{k}/[{(2+k^2)(4+k^2)}]e^{-ikL/2}e^{-\sqrt{4+k^2}L/2}$. Now we can calculate the normalization factor $N_k$ by requiring that \cite{pitaevskii_bose-einstein_2003} $N_k N_{k'}\int dX (u_k u_{k'}^*-v_k v_{k'}^* )=\delta_{k,k'}$. We obtain $N_k=\sqrt{2\varepsilon_k/\widetilde L_s k^2}$, where $\widetilde L_s$ is the dimensionless system size. 
From Eq.~(\ref{eq:Ssetup}) follows the final expression for  $\hat{\psi}_1$ [cf.~Eq.~(\ref{eq:uv})].

We  now consider the second order contribution to the field operator $\hat{\psi}$, described by $\hat\psi_2(X,T)$. Its equation of motion  is obtained from \eq{eq:eom} at order $\alpha^2$. In what follows, we are interested only in the real part of the expectation value $\langle\hat\psi_2\rangle$, since it enters the effective interaction between the impurities at order $\wt G^2$. Introducing the notation
$\psi_2=\text{Re}\langle\hat\psi_2\rangle$, the equation of motion becomes $\mathcal{L}\psi_2(X)=f(X)$, where we define the operator $\mathcal L=-{\partial_X^2/2} -1+3\psi_0^2+L\wt G \delta(X^2- L^2/4)$. Since the source function $f(X)=-2\psi_0\la\hat\psi_1^\da\hat\psi_1\ra-\psi_0\la\hat\psi_1^2\ra$ is time independent, $\psi_2(X)$ also does not depend on time. At zero temperature, $f(X)=-\psi_0\sum_k N_k^2 (2|v_k|^2-u_kv_k^*)$. We point out that the latter sum requires a small-$k$  cutoff, since the summand is divergent at $k\to 0$. However, we emphasize that the final expression for the effective impurity interaction is cutoff independent, as we demonstrate below. The solution for $\psi_2$ can be expressed as $\psi_2(X)=\int dY\, \mathcal G(X,Y)f(Y)$
where $ \mathcal G$ is the Green's function of the operator $ \mathcal L$ and satisfies $\mathcal{L G}(X,Y)=\delta(X-Y)$. Since the source function $f$ can be expanded in powers of $\wt G$ as $f(X)=f_0(X)+\wt G f_1(X)+\mathcal O(\wt G^2)$, we assume that  the Green's function can also be expanded in powers of $\wt G$ as $\mathcal G=\mathcal G_0+\wt G\mathcal G_1+\mathcal O(\wt G^2)$. 
Then $\psi_2$ becomes
\begin{align}\label{seron}
\psi_2(X)={}&\int dY \mathcal G_0(X,Y) f_0(Y)+\wt G\int dY\left[  \mathcal G_0(X,Y) f_1(Y)+\mathcal G_1(X,Y) f_0(Y) \right],
\end{align}
at order $\wt G$. It remains to  determine $\mathcal G_0$ and $\mathcal G_1$. The former satisfies $\mathcal L_0\mathcal G_0(X,Y)=\delta(X-Y)$, where $\mathcal L_0=-\partial_X^2/2+2$ and reads as $\mathcal G_0(X,Y)=e^{-2|X-Y|}/2$.  
$\mathcal{ G}_1(X,Y)$ satisfies $\mathcal L_0\mathcal{ G}_1(X,Y)=-\mathcal L_1\mathcal{ G}_0(X,Y)$, where $\mathcal L_1=L \delta(X^2- L^2/4)+3(\partial_{\wt G}\psi_0^2)|_{\wt G=0}$. We solve this equation and find
\begin{align}
\mathcal{G}_1(X,Y)=-\frac{1}{4}e^{-|L+2X|-|L+2Y|}+\frac{3}{8}\int\dif z e^{-|z-2X|-|z-2Y|-|z+L|}+ (L\to -L),
\end{align}
where $(L\to -L)$ denotes that the previous terms should be evaluated with this replacement. Using Eq.~(\ref{seron}) one can now find $\psi_2$. 

Having determined $\psi_0$, $\hat\psi_1$ and $\psi_2$, we are ready to evaluate the impurity interaction mediated by the Bose gas. At second order in $\wt G$, by making use of the Hellmann-Feynman theorem, we express the effective interaction as
\begin{align}
U=\frac{\mu}{\alpha^2}\wt G\left[ \la \hat \psi^\da \hat\psi \ra |_{X=L/2} -\lim_{L\to \infty}\la \hat\psi^\da \hat\psi \ra |_{X=L/2}\right].
\end{align}
Here the symmetry of the system under the transformation $X\to -X$ enabled us to consider the local density of bosons at the position of only one impurity. Using the perturbative expansion (\ref{eq:wf_exp}), we write the effective interaction  as $U= \frac{\mu}{\alpha^2}[U_0+\alpha^2  U_2+\mathcal O(\alpha^3)]$. We find $U_0= \wt G \big( |\psi_0^2(L/2)|-\lim_{L\to\infty}|\psi_0^2(L/2)|\big)=-\wt G^2 e^{-2L}$. The next order contribution is  $U_2(L)=\mathcal{U}_2(L)-\lim_{L\to \infty}\mathcal{U}_2(L)$, where $\mathcal{U}_2(L)=\wt G \left[\la \hat\psi_1(L/2)^\da\hat\psi_1(L/2)\ra +2\psi_0(L/2)\psi_2(L/2)\right]$. We point out that both contributions in $\mathcal{U}_2(L)$ separately diverge and require an infrared cutoff of the order of the inverse system length. This signals that long wavelength fluctuations destroy the condensate in one dimension in an infinite system. However, the sum of  the two contributions in $\mathcal{U}_2(L)$ is finite and no infrared cutoff is needed. This is expected since $\mathcal{U}_2(L)/\wt G$ is the fluctuation correction of the mean-field boson density. Finally, it remains to express the chemical potential $\mu$ in terms of the boson density. This can be done by inverting the dependence $n(\mu)=\mu\int dX[\psi_0^2+\alpha^2(\la\hat\psi_1^\da\hat\psi_1\ra+2\psi_0\psi_2)+\mathcal O(\alpha^3)]/g\wt L_s$. One obtains $\mu(n)=g n\left(1-\sqrt{\gamma}/\pi\right)$ where $\gamma=mg/\hbar^2 n\ll 1$. Then, summing up all the contributions we obtain Eq.~(\ref{eq:Final}). An alternative derivation of Eq.~(\ref{eq:Final}) based on the density-phase representation of the field operator is given in Appendix.

\section{Discussion and conclusions} The induced impurity interaction (\ref{eq:Final}) is quadratic in $G$ and thus independent of the sign of coupling of the impurities to the liquid. It shows how the classical exponential interaction (\ref{Uexp}) crosses over into the Casimir-like long-range interaction (\ref{Uinf}) as the impurity separation $\ell$ increases. We notice that the full form of Eq.~(\ref{eq:Final}) applies to all distances below $\xi\gamma^{-1/4}$, which is huge for weakly interacting bosons. This limitation arises due to the change of the quasiparticle dispersion at very low momenta where the Bogoliubov spectrum ceases to be valid \cite{imambekov_one-dimensional_2012,pustilnik_low-energy_2014,petkovic_spectrum_2018}. However, its leading linear term properly describes the quasiparticles at small momenta. Therefore, the limiting case of Eq.~(\ref{eq:Final}) given by the dominant term in Eq.~(\ref{Uinf}) holds even at $\ell\to\infty$. We notice that the result (\ref{eq:Final}) can be also obtained using a complementary field-theoretical approach based on the path integral \cite{Reichert_2018}. 

The Casimir-like interaction studied in this work decays much  slower than the Casimir-Polder interaction between the impurities (neutral atoms) that at small distances scales as $\sim 1/\ell^6$ and crosses over into $-\mathcal C\hbar c\alpha_p^2/\pi \ell^7$ at large distances \cite{casimir_influence_1948}. Here $\mathcal C$ denotes a geometric factor of order unity, while $\alpha_p$ denotes the static polarisability of the impurities. Comparing the long range behavior of the two interactions, we find that they are of the same magnitude at $\ell_0/\xi\approx (32 \mathcal C)^{1/4}\sqrt{\gamma\alpha_p n^3  g \sqrt{c}/G \sqrt{v}}$, where $v=\sqrt{g n/m}$ denotes the sound velocity and $c$ is the speed of light. 
Considering, for example, Yb atoms as impurities, $\alpha_p\approx 21$ \AA$^3$.
The typical values \cite{hofferberth_probing_2008}  for a Bose gas of $^{87}\mathrm{Rb}$ are $\gamma=0.005$ and $n=60\, \mu \mathrm{m}^{-1} $, leading to $v=0.32\,\mathrm{cm/s}$ and $\xi=0.24\,\mathrm{\mu m}$. For $G=4g$ we get $\ell_0\approx 0.1 \xi$. Thus at larger distances, the Casimir-like interaction is the dominant one. For impurity separation of the order of the healing length, Eq.~(\ref{eq:Final}) gives the experimentally measurable value of $0.3\,\textrm{kHz}$, while the Casimir-Polder interaction is six orders of magnitude smaller.

To conclude, we have studied the Casimir-like interaction between impurities  in a one-dimensional Bose gas and found the analytic expression (\ref{eq:Final})  for the induced impurity interaction that is valid practically at all distances in the weakly-interacting case. Our work resolves the existing discrepancies in the literature \cite{klein_interaction_2005,recati_casimir_2005,dehkharghani_coalescence_2018,
schecter_phonon-mediated_2014} regarding the form of the induced interaction. In the view of rapidly growing experimental interest in ultracold gases, it is realistic that our result could be tested in the near future. 

\section{Acknowledgments} We thank K. Matveev for useful discussions in the initial stage of this project.

\section{Appendix}

We present here an alternative approach that does not rely on the expansion (\ref{eq:wf_exp}) and demonstrate that it leads to the  same result for the induced interaction between the impurities given by Eq.~(1) in the main text. 
Let us start with the Hamiltonian [Eq.~(4) of the main text] written in the dimensionless form
	\be
		 {\wt H}=\int dX \hat{\wt\Psi}^\da \left[-\dfrac{\partial_X^2}{2}+ \dfrac{1}{2}\hat{\wt\Psi}^\da \hat{\wt\Psi}+ V(X)\right] \hat{\wt\Psi} ,\label{HGC}
	\ee
where $V(X)=L \widetilde G \delta(X^2-L^2/4)$ and $\hat{\wt\Psi}(X,T)=\sqrt{g/\mu} \hat\Psi(X \hbar/\sqrt{\mu m},T\hbar/\mu)$ in terms of $\hat\Psi$ that is defined in the main text. 

In order to treat correctly the phase  of the field operator \cite{girardeau,q.optics,mora_extension_2003}, we need to discretize the space. The   dimensionless lattice spacing is $a$ and is chosen such that it is smaller than all the other length scales characterising  the system, i.e., the healing length and the distance between impurities $L$. The commutation relation reads as
	\be
		[\hat{\wt\Psi}(X,T),\hat{\wt\Psi}^\da(X',T)]=\dfrac{\alpha^2}{a}\delta_{X,X'},
	\ee
where we remind the reader that $\alpha=(\gamma gn/\mu)^{1/4}\ll1$ for a weakly-interacting Bose gas.
The Laplacian operator is defined as
	\be
		\Delta f(X)=\dfrac{f_+(X)+f_-(X)-2f(X)}{a^2},
	\ee
where for notational convenience, $f(X\pm a)=f_\pm(X)$. 
The integral becomes 
	$
		\int dX=\sum_X a .
	$
We now introduce the phase-density representation of the field operators 
	\be
	\hat{\wt\Psi}(X)=e^{-i \tt(X)}\sqrt{\nn(X)},\qquad \hat{\wt\Psi}^\da(X)=\sqrt{\nn(X)}e^{i \tt(X)},
	\ee
where the density and the phase operators satisfy the commutation relation
	\be
		[\nn(X,T),\tt(X',T)]=-\dfrac{i\alpha^2}{a}\delta_{X,X'}\label{com}.
	\ee
The Hamiltonian written in terms of $\nn$ and $\tt$ on the lattice takes the form 
	\begin{align}
		\wt H=&-\dfrac{1}{2a^2}\sum_X a \sqrt{\nn}\left[e^{i(\tt-\tt_+)}\sqrt{\nn_+}+e^{i(\tt-\tt_-)}\sqrt{\nn_-}-2\sqrt{\nn} \right]+\sum_Xa \left[ \nn\left(V(X)-\dfrac{\alpha^2}{2a} \right)+\dfrac{\nn^2}{2}\right], \label{Hdisc}
	\end{align}
where the last term $-(\alpha^2/2)\sum_X \nn$ arises from rewriting the interaction as   $\hat{\wt\Psi}^\da\hat{\wt\Psi}^\da\hat{\wt\Psi}\hat{\wt\Psi}= \hat{\wt\Psi}^\da\hat{\wt\Psi}\hat{\wt\Psi}^\da\hat{\wt\Psi}-(\alpha^2/a)\hat{\wt\Psi}^\da\hat{\wt\Psi}$.
The lattice constant is chosen in such a way that the difference of the phase at neighbouring sites is small. Thus we can expand the exponential functions in \eq{Hdisc} in the difference of the phase at neighbouring sites $\tt-\tt_\pm\sim\epsilon$. At order $\epsilon^0$, one obtains
	\begin{align}
		 H^{(0)}=&\sum_Xa \left[-\dfrac{\sqrt{\nn}\Delta\sqrt{\nn}}{2} +\nn\left(V(X)-\dfrac{\alpha^2}{2a} \right)+\dfrac{\nn^2}{2}\right].
	\end{align}
At linear order in $\epsilon$, one has
\begin{align}
		 H^{(1)}=&-\dfrac{i}{2a^2}\sum_Xa \sqrt{\nn}\left[(\tt-\tt_+)\sqrt{\nn_+}+(\tt-\tt_-)\sqrt{\nn_-} \right]
			=-\dfrac{\alpha^2}{4a}\sum_Xa \dfrac{\Delta\sqrt{\nn}}{\sqrt{\nn}}-\dfrac{\alpha^2}{2a^2}\sum_X 1,
	\end{align}
where the commutation relation~(\ref{com}) was used in order to eliminate $\tt$. 
At order $\epsilon^2$, one finds
	\begin{align}
		 H^{(2)}={}&\dfrac{1}{4a^2}\sum_Xa \sqrt{\nn}\left[(\tt-\tt_+)^2\sqrt{\nn_+}+(\tt-\tt_-)^2\sqrt{\nn_-} \right]\notag  \\
			={}&\dfrac{1}{4}\sum_Xa\left[ \sqrt{\nn}\Delta(\tt^2\sqrt{\nn})+\sqrt{\nn}\tt^2\Delta\sqrt{\nn} -2\sqrt{\nn}\tt\Delta(\tt\sqrt{\nn})\right].
	\end{align}
Higher order terms proportional to the $\epsilon^{n}$, $n\geq 3$, will vanish in the continuum, i.e., in $a\to 0$ limit since $\epsilon\sim \tt-\tt_\pm  \approx a \partial_X\hat\theta$. Let us now derive the equations of motion for $\nn$ and $\tt$. These are given by $\alpha^2\partial_T \nn =i[\wt H,\nn]$ and  $\alpha^2\partial_T \tt =i[\wt H,\tt]$ where here $\wt H\simeq  H^{(0)}+ H^{(1)}+ H^{(2)}$. One obtains
\begin{align}
\partial_T \nn={}&\dfrac{1}{2}\left\{ \sqrt{\nn}\left[\Delta(\tt\sqrt{\nn})-\tt\Delta \sqrt{\nn} \right]+\text{h.c.}\right\},\label{eomnd}\\
\partial_T \tt=&-\dfrac{\Delta\sqrt{\nn}}{2\sqrt{\nn}}+\dfrac{1}{8}\left\{\dfrac{1}{\sqrt{\nn}}\left[\Delta(\tt^2\sqrt{\nn})+\tt^2\Delta\sqrt{\nn}-2\tt\Delta(\tt\sqrt{\nn}) \right]+\text{h.c.} \right\}+\nn+V(X)\notag\\ &-\dfrac{\alpha^2}{2a}\left\{1+\dfrac{1}{4\sqrt{\nn}}\left[\Delta\left(\dfrac{1}{\sqrt{\nn}}\right)-\dfrac{\Delta\sqrt{\nn}}{\nn} \right] \right\}.\label{eomtd}
\end{align}
We can now consider the limit $a \to 0$ and obtain the following equations valid in the continuum
\begin{gather}
\partial_T \nn=(\tt'\nn)',\label{eomn}\\
\partial_T \tt=-\dfrac{\nn''}{4\nn}+\dfrac{(\nn')^2}{8\nn^2}+\dfrac{(\tt')^2}{2}+\nn+V(X)-\dfrac{\alpha^2\delta(0)}{2}\left[ 1-\dfrac{(\ln\nn)''}{4\nn}\right]\label{eomt}.
\end{gather}
Here the prime denotes the derivative with respect to $X$. We used the fact that $\alpha^2\delta_{X,X'}/a\to\alpha^2\delta(X-X')$ in the limit $a\to 0$.

We now solve Eqs.~(\ref{eomn}) and (\ref{eomt}). The phase can be written as $\tt=T+\delta\tt$, where $T$ is dimensionless form of the $\mu t/\hbar$ factor and $\delta\tt$ denotes the fluctuating part. In the case of weakly-interacting bosons, we express the density  as  $\nn=n_0+\delta\nn$, where the density fluctuation contribution $\delta \nn(X,T)$ around the background mean-field density $n_0(X)$ is small. Minimizing the Hamiltonian (\ref{HGC}) in the absence of fluctuations with respect to $n_0$ gives the equation
\be
	-\dfrac{n_0''}{4}+\dfrac{(n_0')^2}{8n_0}+n_0^2+[V(X)-1]n_0=0.\label{eom0}
\ee
Alternatively, the latter equation can be obtained from Eq.~(\ref{eomt}) in the absence of fluctuations ($\alpha\to 0$).
Notice that Eq.~(\ref{eom0}) is equivalent to the Gross-Pitaevskii (\ref{Psi_0}) equation for $\psi_0$ obtained in the main text, provided  
\begin{align}
\psi_0=\sqrt{n_0}.\label{zzzz}
\end{align}
At leading order in quantum fluctuations,  Eq.~(\ref{eomt}) becomes 
	\be
		\partial_T\delta \tt=\dfrac{(\delta\tt')^2}{2}+\mathcal O\big(\delta\nn/n_0 ,\alpha^2\big),
	\ee
using Eq.~(\ref{eom0}). 
Since in equilibrium $\la \partial_T \delta\tt\ra=0$ \cite{mora_extension_2003}, the previous equation implies the smallness of  $\delta\tt'$. From the commutation relation follows
	\begin{gather}
\label{zzzz1}	\delta\nn=\alpha\nn_1+\alpha^2\nn_2+\ldots,\\
	\delta\tt'=\alpha \tt_1'+\alpha^2\tt_2'+\ldots.
	\end{gather}
The equations of motion for $\tt_1$, $\nn_1$, $\tt_2$, and $\nn_2$  are obtained by expanding Eqs.~(\ref{eomn}-\ref{eomt}) in $\alpha$. At linear order in $\alpha$, we obtain
	\begin{gather}
		\partial_T \nn_1 = \big(\tt_1'n_0 \big)', \label{eomn1}\\
		\partial_T \tt_1=-\dfrac{\nn_1''}{4n_0}+\dfrac{n_0'\nn_1'}{4n_0^2}+\left[2n_0+V(X)-1-\dfrac{\big( n_0'\big)^2}{8n_0^2} \right]\dfrac{\nn_1}{n_0},\label{eomt1}
	\end{gather}
by making use of Eq.~(\ref{eom0}).
From these two equations of motion and Eq.~(\ref{eom0}) follows that the linear combination ${\nn_1}/{2\sqrt{n_0}}-i\sqrt{n_0}\tt_1$ satisfies the same equation of motion as $\hat \psi_1$ in the main text. Remarkably, we can thus identify \cite{mora_extension_2003}
	\be
		\hat\psi_1=\dfrac{\nn_1}{2\sqrt{n_0}}-i\sqrt{n_0}\tt_1,\quad \hat\psi_1^\da=\dfrac{\nn_1}{2\sqrt{n_0}}+i\sqrt{n_0}\tt_1.\label{psi1}
	\ee
Using the solution for $\hat\psi_1$  from the manuscript, one can now obtain   $\hat\theta_1$ and show that the phase fluctuations destroy the condensate in the thermodynamic limit, i.e., $\la e^{-i\alpha \hat\theta_1}\ra=0 $. Note that  $ [\nn_1(X,T),\tt_1(Y,T)]=-i\delta(X-Y)$.

The equations of motion of $\tt_2$ and $\nn_2$ are obtained by expanding Eqs.~(\ref{eomn}-\ref{eomt}) at second order in $\alpha$. They read as
	\begin{gather}
		\partial_T \nn_2 = \big(\tt_2'n_0 \big)'+ \big(\tt_1'\nn_1\big)', \label{eomn2}\\
		\begin{split}\partial_T \tt_2=&-\dfrac{\nn_2''}{4n_0}+\dfrac{n_0'\nn_2'}{4n_0^2}+\left[2n_0+V(X)-1-\dfrac{( n_0')^2}{8n_0^2}\right]\dfrac{\nn_2}{n_0}+\dfrac{\nn_1\nn_1''}{4n_0^2}+\dfrac{(\nn_1')^2}{8n_0^2}-\dfrac{n_0'(\nn_1^2)'}{4n_0^3}\\
		&-\dfrac{\nn_1^2}{n_0^2}\left[n_0+V(X)-1-\dfrac{(n_0')^2}{4n_0^2} \right]+\dfrac{(\tt_1')^2}{2}-\dfrac{\delta(0)}{2}\left[1-\dfrac{(\ln n_0)''}{4n_0}\right].
	\end{split}\label{eomt2}	
	\end{gather}
We used Eq.~(\ref{eom0}) in order to eliminate terms in $n_0''$.

We can derive, using Eqs.~(\ref{eom0}), (\ref{eomn1}), (\ref{eomt1}), (\ref{eomn2}), and (\ref{eomt2}), the following equations
	\begin{align}\label{eq1}
		&-\partial_T\left(\sqrt{n_0} \la\tt_2\ra+\dfrac{1}{4\sqrt{n_0}}\la\nn_1\tt_1+\tt_1\nn_1\ra\right)\notag\\& +\left[-\dfrac{\partial_X^2}{2}+3n_0+V(X)-1\right] \dfrac{\la\nn_2\ra-\frac{\la\nn_1^2\ra}{4n_0}-n_0\la\tt_1^2\ra+\frac{\delta(0)}{2}}{2\sqrt{n_0}}=\dfrac{\la \hat{f}+\hat{f}^\da\ra}{2},\\
		&\partial_T\left( \dfrac{\la\nn_2\ra-\frac{\la\nn_1^2\ra}{4n_0}-n_0\la\tt_1^2\ra+\frac{\delta(0)}{2}}{2\sqrt{n_0}}\right)\notag\\&+\left[-\dfrac{\partial_X^2}{2}+n_0+V(X)-1\right]\left(\sqrt{n_0} \la\tt_2\ra+\dfrac{1}{4\sqrt{n_0}}\la\nn_1\tt_1+\tt_1\nn_1\ra\right)=i\dfrac{\la \hat{f}-\hat{f}^\da\ra}{2},
	\end{align}
where 
	\begin{align}
		\hat{f}=&-\dfrac{3\nn_1^2}{4\sqrt{n_0}}-n_0^{3/2}\tt_1^2-\dfrac{i\sqrt{n_0}}{2}\tt_1\nn_1+\dfrac{3i\sqrt{n_0}}{2}\nn_1\tt_1 \label{f1}\\
			=&-{\psi_0}\big[2 \hat\psi_1^\da\hat\psi_1+(\hat\psi_1)^2\big].\label{f2}
	\end{align}
We used that $\sqrt{n_0}=\psi_0$ and Eq.~(\ref{psi1}) in Eq.~(\ref{f2}). 
By identifying
	\begin{gather}
		 \dfrac{\la\nn_2\ra-\frac{\la\nn_1^2\ra}{4n_0}-n_0\la\tt_1^2\ra+\frac{\delta(0)}{2}}{2\sqrt{n_0}}= \dfrac{\la\nn_2\ra-\la\hat\psi_1 ^\da\hat\psi_1\ra}{2\psi_0}=\text{Re} \la \hat{\psi}_2\ra,\label{re}\\
		-\sqrt{n_0} \la\tt_2\ra-\dfrac{1}{4\sqrt{n_0}}\la\nn_1\tt_1+\tt_1\nn_1\ra=-\psi_0 \la\tt_2\ra +\dfrac{i}{4\psi_0}\la (\hat\psi_1^\da)^2-(\hat\psi_1)^2 \ra =\text{Im} \la\hat{\psi}_2\ra ,\label{im}
	\end{gather}
the equations of motion for $\text{Re} \la\hat\psi_2\ra$ and $\text{Im} \la\hat\psi_2\ra$ read as
\be
	L |\la\hat\psi_2\ra\ra=|\la\hat{f}\ra\ra,\label{eompsi2}
\ee
where $|\la a\ra\ra=(\la a\ra,\la a^\da\ra)^T$ and
	\be
		 L=\left(\begin{matrix}
		-i\partial_T-\dfrac{\partial_X^2}{2}+2\psi_0^2+V(X)-1&\psi_0^2\\
		\psi_0^2&i\partial_T-\dfrac{\partial_X^2}{2}+2\psi_0^2+V(X)-1
		\end{matrix}\right).
	\ee
Using the solution for $\hat\psi_1$ from the manuscript one sees that $\la (\hat\psi_1)^2 \ra$ is real. This  implies $\la \nn_1\tt_1+\tt_1\nn_1\ra=0$ [see Eq.~(\ref{psi1})]. Moreover, from the equation of motion of $\la\hat\psi_2\ra$ [Eq.~(\ref{eompsi2})] follows that $\la\hat\psi_2\ra$ is time independent since the source term $|\la\hat f\ra\ra$ is time independent. Thus from Eq.~(\ref{im}) we get $\partial_T \la\tt_2\ra=0$. Using the fact that $\la\hat f\ra=\la\hat f^\da\ra$, Eq.~(\ref{eq1}) becomes 
\begin{align}
\left(-\dfrac{\partial_X^2}{2}+3\psi_0^2+V(X)-1\right)\text{Re} \la \hat{\psi}_2\ra=\la\hat f\ra. 
\end{align} 
This is the equation for $\psi_2=\text{Re} \la \hat{\psi}_2\ra$ given in the manuscript with $f=\la \hat f\ra$. 

For the evaluation of the induced interaction between the impurities, as shown in the manuscript, the observable of interest is the density. To order $\alpha^2$, using Eqs.~(\ref{zzzz}), (\ref{zzzz1}), (\ref{psi1}), and (\ref{re}), the expectation value of the density is $\la \hat n\ra=n_0+\alpha \la\nn_1\ra+\alpha^2 \la\nn_2\ra=\psi_0^2+\alpha^2(2\psi_0\text{Re} \la \hat{\psi}_2\ra+\la\hat\psi_1 ^\da\hat\psi_1\ra)$. This is the expression obtained in the main text using a complementary approach. Thus Eq.~(1) of the main text does not require the expansion (\ref{eq:wf_exp}) and can be derived using the density-phase representation, as demonstrated in this Appendix.


%

\end{document}